\documentclass{aa}  

\usepackage{graphicx}
\usepackage{txfonts}
\usepackage{xcolor}
\usepackage{tabularx}
\usepackage{appendix,natbib}
\usepackage{amsmath}
\usepackage{subfigure}
\usepackage[breaklinks=true]{hyperref} 

\begin{document}

   \title{Dark matter fraction in $z\sim 1$ star-forming galaxies}


   \author{Gauri Sharma, 
          \inst{1,2,3} \fnmsep\thanks{Contact: gsharma@sissa.it}
          Paolo Salucci \inst{1,2,3}
          \and
         Glenn van de Ven\inst{4}
          }

   \institute{SISSA International School for Advanced Studies, Via Bonomea 265, I-34136 Trieste, Italy
         \and
           GSKY, INFN-Sezione di Trieste, via Valerio 2, I-34127 Trieste, Italy
         \and
           IFPU Institute for Fundamental Physics of the Universe, Via Beirut, 2, 34151 Trieste, Italy
         \and 
           Department of Astrophysics, University of Vienna, T\"urkenschanzstrasse 17, 1180 Wien, Austria               
             }

   \date{Received 11/02/2021; accepted 27/05/2021}

 
  \abstract
   {The study of dark matter (DM) across cosmic timescales is essential for understanding galaxy formation and evolution. Recent observational studies show that further back in time ($z >0.5$), rotation-supported, star-forming galaxies (SFGs) begin to appear to be DM deficient compared to local SFGs.}
   {We present an observational study of the DM fraction in 225 rotation-supported, SFGs at $z\approx 0.9;$ these SFGs have stellar masses in the range $ 9.0 \leq log(M_* \ \mathrm{M_\odot}) \leq 11.0$ and star formation rates $0.49 \leq log \left(SFR \ \mathrm{[M_{\odot}\ yr^{-1}]} \right) \leq 1.77$.}
   {We studied a subsample of the KMOS Redshift One Spectroscopic Survey (KROSS) studied by \citet{GS20}. The stellar masses ($M_*$) of these objects were previously estimated using mass-to-light ratios derived from fitting the spectral energy distribution of the galaxies. Star formation rates were derived from the H$_\alpha$ luminosities. In this paper, we determined the total gas masses ($M_{gas}$) by the scaling relations of molecular and atomic gas \citep[][respectively] {Tacconi2018, Lagos2011}. We derived the dynamical masses ($M_{dyn}$) from the rotation curves (RCs) at different scale lengths (effective radius: $R_e$, $\sim 2 \ R_e$ and $\sim 3 \ R_e$) and we then calculated the DM fractions ($f_{ DM }=1-M_{bar}/M_{dyn}$) at these radii.}
   {We report that at $z\sim 1$ only a small fraction ($\sim 5\%$) of our sample has a low ($< 20\%$) DM fraction within $\sim$ 2-3 $R_e$. The majority ($> 72\%$) of SFGs in our sample have outer disks ($\sim 5-10$ kpc) dominated by DM, which agrees with local SFGs. Moreover, we find a large scatter in the fraction of DM at a given stellar mass (or circular velocity) with respect to local SFGs, suggesting that galaxies at $z \sim 1$ span a wide range of stages in the formation of stellar disks and have diverse DM halo properties coupled with baryons.}

   \keywords{galaxies: kinematics and dynamics;-- galaxies: late-type, disk-type and rotation dominated; -- galaxies: evolution; -- galaxies: dark matter halo;-- cosmology: nature of dark matter
               }

   \maketitle
%

\section{Introduction}
It has been known for decades that the rotation curve (RC) of a galaxy can be used as a proxy for the enclosed mass and its underlying distribution (\citealt{Sofue2001} and  \citealt{PS2019}). In the local Universe, we see the RCs of late-type galaxies\footnote{In the local Universe, most of the late-type galaxies are spiral systems hosting star-forming disks and are rotation-supported ($v/\sigma >1 $).} rising steeply in the inner regions and then flattening far beyond the disk edge\footnote{The disk edge is defined as $3.2 \ R_D$ ( $=$ $1.89 \ R_e$), where the stellar surface luminosity $\propto \exp(-r/R_D)$} \citep[][and references therein]{rubin1980, PS1996, McGaugh2016}. The steep rise in the inner regions is caused by the combination of a baryon-dominated disk plus a cored halo (\citealt{PS1996} and \citealt{PS1988}), while the flattening in the outer regions of the stellar disk and beyond implies dark matter (DM) dominance \citep[][and references therein]{PS1996, PS1988, Kassin2006, Martinsson2013, Courteau2015}. Recently, RC studies have been extended to higher redshifts (e.g., \citealt{ETD16} and \citealt{Genzel2017}). However, as a result of observational challenges (e.g., a limited spatial resolution and moderate signal-to-noise ratio), the interpretation of the results remains ambiguous and  still needs to converge to more concrete conclusions.

Recently, \citet{Genzel2017} and \citet{Lang2017} analyzed the RCs of star-forming galaxies (SFGs) at high-$z$ and found a declining behavior with increasing radius; such behavior is only seen in very massive local SFGs, while the RCs of normal SFGs are remarkably flat and rarely decline (e.g., \citealt[][]{rubin1980} and \citealt{PS1996}).  \citet{Genzel2017} studied the individual RCs of six massive ($log(M_* \ \mathrm{[M_\odot]})=10.6-11.1$) SFGs at redshift $0.9\leq z\leq 2.4 $ in more detail. They showed declining RCs in two cases:  first, when the individual RCs were normalized at $R_{max}$, where the amplitude of the rotational velocity is maximal; and second when the binned average of the six individual galaxies were normalized at the effective radius ($R_e$). Based on these results, \citet{Genzel2017} concluded that the fraction of DM within $R_e$ is below $20\%$. Later, these results were confirmed by \citet{Lang2017}, who obtained the stacked normalized RCs of 101 SFGs at $0.6 \lesssim z \lesssim 2.2$ with stellar masses $9.3 \lesssim log(M_* \ \mathrm{[M_\odot]}) \lesssim 11.5$, where the normalization of RCs was performed at a ``turnover radii', where $R_{turn} \sim 1.65 \ R_e$ (for details see \citealt{Lang2017}). Both studies suggest that the declining behavior of RCs can be explained by a combination of high baryon fraction and extensive pressure support. Some other high-$z$ studies of late-type and early-type galaxies also report similar low DM fractions within the effective radii \citep[e.g.,][]{Burkert2016, Wuyts2016, Price2016, Ubler2018}.

On the other hand, \citet{AT2019} studied the shape of RCs in $\approx 1500$ SFGs at high-$z$ ($0.6\lesssim z \lesssim 2.2$) that have stellar masses $8.5 \lesssim log(M_* \ \mathrm{[M_\odot]}) \lesssim 11.7 $. These authors used a similar stacking approach as \citet{Lang2017} and confirmed that the RCs are similar to those in \citet{Lang2017} (i.e., declining) if they are normalized at turnover radii. However, if instead the normalization was performed at $3 \times R_D \ (\approx 1.77 \ R_e)$ they found flat or rising RCs. Moreover, \citet{AT2019} reported a more than $50\%$ DM fraction within $3.5\ R_e$, which is similar to local star-forming disk galaxies \citep{PS1996, Martinsson2013, Courteau2015}. Finally, another study by \citet{Drew2018}  analyzed a massive star-bursting galaxy at $z=1.55$, which showed strong evidence of a flat RC between $6-14$ kpc with a $44\%$ DM fraction within the effective radius.

In order to resolve the serious issues raised by the previous literature on high-$z$ RCs, we recently studied the intrinsic shape of RCs \citep{GS20} by means of 256 SFGs that have stellar masses $8.83 \leq \log (M_{*} \ \mathrm{[M_{\odot}]}  \leq 11.32$) at $0.7\lesssim z \lesssim 1.04$. In particular, we exploited KROSS data \citep{stott2016, H17} by employing different techniques than used in \citet{Genzel2017, Lang2017} and \citet{AT2019b}. In brief, the key differences between our findings and previous studies lie in the modeling of the kinematics and the consideration of pressure support. We performed 3D forward modeling of data cubes using the $^{3D}$Barolo code \citep{ETD15}, which improves the quality of resulting kinematics with respect to 2D kinematic modeling approaches. For pressure support, we employed the pressure gradient correction (PGC) method, which makes use of the full information available in the data cube and therefore gives better results than pressure support correction assuming isotropic and constant velocity dispersion; details are available in \citet[][appendix]{GS20} and \citet{kretschmer2020}. Our results confirm that the outer ($5\lesssim R \lesssim 20$ kpc) RCs of $z\sim 1$ galaxies are flattish up to the last observation point and are very similar to the RCs of local star-forming disk galaxies. Noticeably, the inner ($R \lesssim 5$ kpc) RCs of local SFGs are not flat; in this respect only future observations at high-$z$ can show whether the similarity entirely remains or not.

In this paper we address the DM fraction in $z \sim 1$ SFGs using the individual RCs derived in \citet{GS20}. The article is organized as follows: Section[\ref{sec:data}] contains a brief discussion of the data used; in Section[\ref{sec:analysis}] we compute and analyze the stellar, gaseous, and dynamical masses of the sample. In Section[\ref{sec:results}] we present the individual and averaged DM fractions. In the Section[\ref{sec:discussion}] we discuss and compare our results with previous studies, and finally we critically summarize our findings in  Section[\ref{sec:summary}]. Throughout the analysis we assumed a flat $\Lambda$CDM cosmology with $\Omega_{m,0} =0.27$, $\Omega_{\Lambda,0}=0.73$, and $H_0=70 \ \mathrm{km \ s^{-1}}$.

\section{Data}
\label{sec:data}
We utilize the KROSS dataset previously studied by \citet{stott2016, H17, AT2019, AT2019b} and \citet{HLJ17}. \footnote{Data are publicly available on the KROSS website http://astro.dur.ac.uk/KROSS/data.html.} In particular, we studied a subsample of 344 KROSS SFGs in \citet[][]{GS20}, where the kinematics of these objects were rederived from the original KROSS data cubes via the $^{3D}$Barolo code (\citealt{ETD15} and \citealt{ETD16}). This code accounts for the beam-smearing correction in 3D space and provides the moment maps, stellar surface brightness profile,   RC, and dispersion curve (DC) along with the kinematic models. In addition, our RCs were corrected for pressure gradients that are likely to affect the kinematics of high-$z$ galaxies. In the end, we analyzed only 256 rotation-dominated Quality 1 and 2 objects out of 344 KROSS objects, which we refer to as the Q12 sample;  \citet{GS20} gives details on the selection criteria and the dataset. 

In this work, we use the Q12 sample to derive the DM fraction at various scale lengths (characteristic radii)\footnote{In general, the scale lengths are associated with various quantities that decrease exponentially such as the surface brightness.} ; those emerge from the local star-forming disk galaxies (given in, \citealt[][]{Freeman} and \citealt{PS1996}). According to which, the effective radius (half-light radius) $R_e$ of a galaxy is defined to encompass the $50\%$ of total integrated light. For an exponential surface brightness distribution $I(r) \propto \exp (-r/R)$, $R$ can be one of the various scale lengths, each of which is  proportional to the stellar Freeman disk length $R_D$\footnote{$R_D$ is a typical characteristic radius of local star-forming disk galaxies.} \citep{Freeman}; for example, $R_e = 1.69 \ R_D$. Under the same assumption, the optical radius $R_{opt}$ containing 83\% of the integrated light becomes $R_{opt} = 3.2 \ R_D$. In the Q12 sample, the median effective radius is $\sim 3.1 \ \mathrm{kpc}$ and the median optical radius is $\sim 5.9 \ \mathrm{kpc}$. This means that the effective radius of the majority of the sample falls below the resolution limit ($\sim 4.0 $ kpc), since the median seeing is $0.5 \arcsec $ at $z\sim 0.9$, while the optical radius is of the same order. Therefore, to be conservative and trace the DM fraction to the furthest point where we have data, we define the outer radius $R_{out} = 5 \ R_D$, which stays above the resolution limit in majority ($\approx 99\%$) of the sample.\footnote{Scale lengths in terms of the effective radius: $R_D = 0.59 \ R_e$; $R_{opt} = 1.89 \ R_e$; $R_{out}=2.95 \ R_e$}

Furthermore, for the accuracy and quality of the sample, we remove those galaxies that have $R_{out}< 3.5$ kpc (below the resolution limit), $z< 0.65$, and $M_* < 10^9 M_\odot$;\footnote{To derive the gas mass, we used scaling relations that hold for $9.0 \leq log(M_\odot) \leq 11.8$; therefore, we excluded one galaxy ('U-HiZ\_z1\_201') with $log(M_\odot) <9.0$.} in total we removed 31 galaxies. Our final sample contains 225 galaxies that have the inclination range $25^{\circ} < \theta_i \leq 75^{\circ}$, the redshift range $0.76 \leq z \leq 1.04$, the effective radii $0.08 \leq \log (R_{e} \ \mathrm{[kpc]}) \leq 0.89$, and circular velocities $1.45 \leq \log (V_{out} \ \mathrm{[km \ s^{-1}]}) \leq 2.83$, where $V_{out}$ is calculated at $R_{out}$.

\section{Analysis}
\label{sec:analysis}
In this section, we discuss the determination of stellar, gas, and dynamical masses. 
\subsection{Stellar masses}              
\label{sec:Mstar}

We adopt the stellar masses given by \citet{H17}. These stellar masses were calculated using a fixed mass-to-light ratio following $M_* = \Upsilon_H \times 10^{-0.4\times(M_H - 4.71)}$, where $\Upsilon_H$ and $M_H$ are the mass-to-light ratio and absolute magnitude in the H band (rest frame), respectively. Here $\Upsilon_H=0.2$, which is the median value of our sample obtained using the {\sc Hyperz} \citep{Bolzonella2000} spectral energy distribution (SED) fitting tool; this tooluses a suite of spectral templates from \citet{Bruzual2003} and optical to near-infrared (NIR) photometry (U, B, V, R, I, J, H, and K) \footnote{In some cases, mid-infrared IRAC bands were also used; \citet{H17} provide details.}.

The stellar masses calculated from a fixed mass-to-light ratio are slightly different from other studies of the same KROSS sample (e.g., \citealt[][]{stott2016} and \citealt{AT2019}), in which different SED fitting procedures were used. However, the absolute median difference between the previous studies is about $\pm 0.2 \ dex$; see Appendix Figure[\ref{fig:Ms-comparison}]. Therefore, we assigned a homogeneous uncertainty of $\pm 0.2 \ dex$ on $M_*$, which is motivated by the aforementioned studies, as well as accounts for the typical uncertainty owing to low and high signal-to-noise photometry. Our sample covers the stellar mass range $9.0 \leq \log (M_{*} \ \mathrm{[M_{\odot}]} ) < 11.0$. We note that we do not have very massive ($\log (M_{*} \mathrm{[M_{\odot}]} ) >11.0$) galaxies in our sample.

\subsection{Molecular gas masses}
\label{sec:MH2}
To estimate the molecular gas mass ($M_{H2}$) of our galaxies, we adopt the relation given by \cite{Tacconi2018}. These authors used a large sample of 1309 SFGs in the redshift range $z = 0-4.4$ with stellar masses $log \left(M_*\left[M_\odot \right] \right)=9.0-11.9$ and star formation rates (SFRs) $10^{-2}-10^{2} M_\odot \ yr^{-1}$. We stress that \cite{Tacconi2018} combined the three different methods to determine the molecular gas mass, each of which is related to one of the following: 1) CO line flux, 2) far-infrared dust SED, and 3) 1 mm dust photometry. Based on their composite findings, they established a single multidimensional unified scaling relation given by
\begin{multline}
\label{eq:H2gas-frac}
log(M_{H2}) = [A + B\times (\log (1+z)-F)^2 +C \times log(\delta MS)\\ + D \times (\log M_* - 10.7)] + log(M_*),
\end{multline}
where, $A,\  B,\  C, \ D,$ and $F$ are proportionality constants reported in Table(3b) of \citealt{Tacconi2018}, $MS$ is the main sequence (MS) relation of SFGs \citep[see,][]{speagle14}, and $\delta MS$ is the offset from $MS$ line. In detail, $\delta MS = sSFR/sSFR(MS; \ z, \ M_*)$, where $sSFR$ is total specific SFR that is computed as $sSFR = (SFR_{H\alpha}^{int})/M_*$, where $SFR_{H\alpha}^{int}$ is the dust reddening-corrected $H_\alpha$ SFR. In this work the quantity $SFR_{H\alpha}^{int}$ is derived by following the procedure of \citet{stott2016}; for details see Appendix[\ref{sec:SFR}]. The quantity $sSFR(MS; \ z, \ M_*)$ is the specific SFR defined by \citet{speagle14} and can be computed as follows:
\begin{align*}
\log (sSFR(MS; \ z, \ M_*)) & = (-0.16-0.026\times t_c) \\
& \quad \times (\log M_{*} +0.025)\\
& \quad -  (6.51-0.11 \times t_c) + 9,
\end{align*}
where $t_c$ is the cosmic time, given by
\begin{align*}
\log (t_c \ [Gyr]) & = 1.143 - 1.026 \times \log (1+z)\\
& \quad - 0.599 \times \log^2 (1+z)\\
& \quad + 0.528 \times \log^3 (1+z).
\end{align*}
The full derivations of $M_{H2}$ and $MS$ of galaxies at different redshifts are explained in greater detail in \citealt{Tacconi2018} and \citet{speagle14}, respectively.

The galaxies in our sample follow the relation between star formation rate and stellar mass given by \citet{speagle14}, the so-called main sequence, with the following range of the relevant quantities: redshifts $0.77 \leq z \leq 1.04$, stellar masses $9.0 \leq log \left(M_{*} \ \mathrm{[M_{\odot}]} \right) \leq 10.97$, and SFRs $0.49 \leq log \left(SFR \ \mathrm{[M_{\odot}\ yr^{-1}]} \right) \leq 1.77$ (see Appendix Figure [\ref{fig:SFR}]). Therefore, using the equation [\ref{eq:H2gas-frac}], we estimated the molecular gas mass of our galaxies.

\begin{figure*}
        \begin{center}
                \includegraphics[width=\columnwidth]{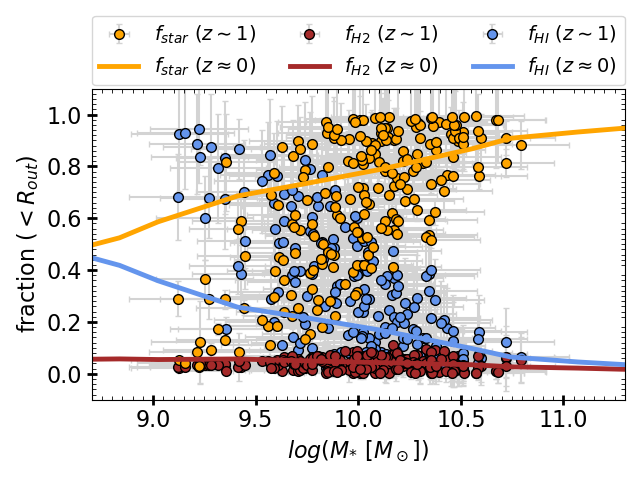}
                \includegraphics[width=\columnwidth]{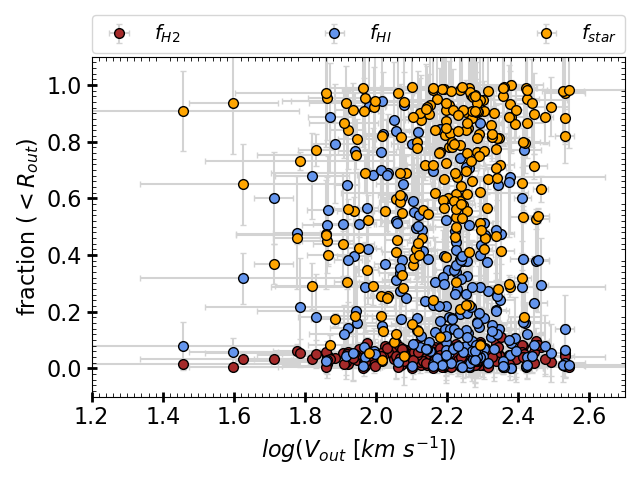}
                \caption{Stellar and gas mass fraction of our sample as a function of stellar mass and circular velocity (left and right panel, respectively) within the outer radius ($R_{out}$; i.e., visible region). The color code in both panels is the same and given as follows: the brown filled  circles represent the molecular gas mass fraction ($f_{H2}=M_{H2}(<R_{out})/M_{bar}(<R_{out})$), the orange filled circles indicate the star mass fraction ($f_{star}=M_{*}(<R_{out})/M_{bar}(<R_{out})$), and the blue filled circles represent the atomic gas mass fraction ($f_{ HI }= M_{ HI }(<R_{out})/M_{bar}(<R_{out})$). A comparison study of local late-type galaxies from \citet{Calette2018} is drawn by solid lines; the color coding is same as the high-$z$ objects.}
                \label{fig:MsVc-gasfrac}
        \end{center}
\end{figure*}

\begin{figure*}
        \begin{center}
                \includegraphics[width=\columnwidth]{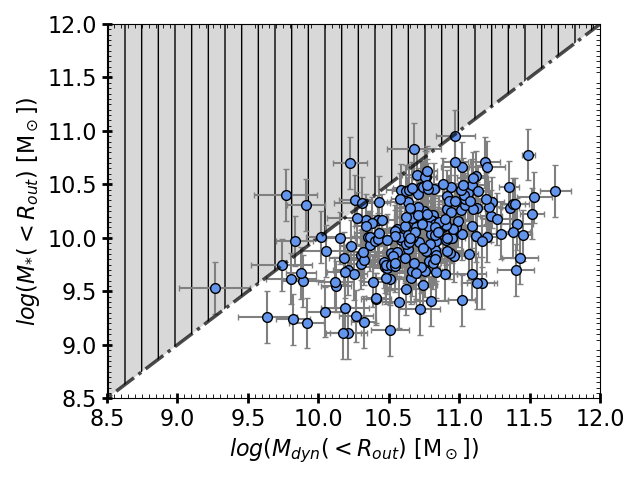}
                \includegraphics[width=\columnwidth]{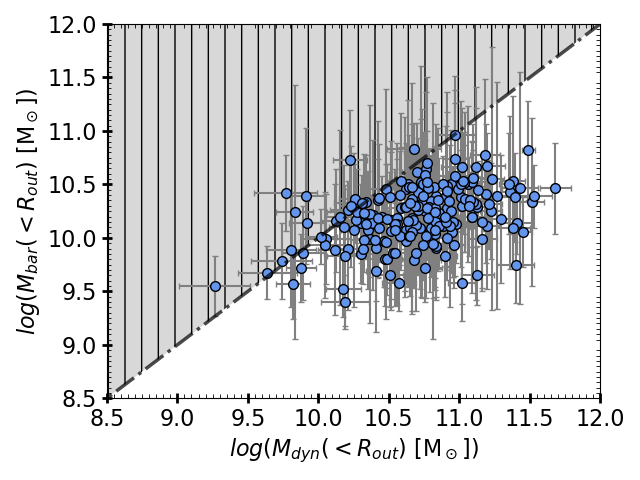}
                \caption{Stellar and baryonic masses as a function of dynamical mass (left and right panels, respectively), computed within $R_{out}$. The notations and color codes are the same in both panels and are given as follows: the blue filled circles represent the data, the dot-dashed black line shows the one-to-one plane, and the hatched-shaded gray area represents the forbidden region. }
                \label{fig:Mdyn-Ms-Mbar}
        \end{center}
\end{figure*}

\subsection{Atomic gas masses}
\label{sec:MHI}
Observation of atomic hydrogen (HI) is challenging and can only be done in the radio wavelength by mapping the 21 cm fine-structure emission line produced by the spin-flip transition of the electron in the H atom. Currently, we cannot observe the 21 cm line at high-$z$, but with next-generation radio telescopes (e.g., ASKAP: \citealt{Johnston2008, McConnell2016} and \citealt{Bonaldi2021}) this should become possible in the near future. At present, the HI gas mass is typically inferred through the spectra of quasars \citep[][and reference therein] {Peroux2003, Prochaska2005, Rao2006, Guimar2009, Noterdaeme2012, Krogager2013, Moller2018}, and scaling relations based on these observations and simulations. In our work, we use the scaling relation given by \citet{Lagos2011} for the atomic gas mass as follows:
\begin{equation}
\label{eq:HIgas-frac}
M_{HI} = \frac{M_{H2}}{0.01} \ \Big(\frac{10^{10} \ h^{-1} \ M_\odot}{M_*} \Big)^{0.8} \ (1+z)^{-3.3}
.\end{equation}
The above relation agrees well with the global density of atomic and molecular hydrogen derived from damped Ly-$\alpha$ observations \citep{Peroux2003, Rao2006, Noterdaeme2009} and predicts the H2 and HI mass function in the local Universe \citep{Zwaan2005, Martin2010}. Moreover, the Equation [\ref{eq:HIgas-frac}] agrees very well with the scaling relation provided by a recent study of \citet{Calette2018} and DustPedia late-type galaxies \citep{Casasola2020}. Therefore, we use this relation to determine the HI masses of our sample. However, we emphasize that the $M_{ HI }$ computed using Equation [\ref{eq:HIgas-frac}] gives the mass of the atomic gas, which corresponds to the scale of the molecular gas. In the outskirts of galaxies, $M_{ HI }$ is most likely different. 
  
\subsection{Baryonic mass estimates}
\label{sec:Mbar}
In the section[\ref{sec:Mstar}, \ref{sec:MH2}, and \ref{sec:MHI}], we derived the total stellar, molecular, and atomic mass content of our galaxies,  which lead us to estimate the total baryonic mass as follows:
\begin{equation}
\label{eq:Mbar}
M_{bar} = M_* + \ M_{H_{2}} + 1.33 M_{HI}
,\end{equation}
where factor 1.33 accounts for the helium abundance. As a result, the total gas and stellar mass fraction can be written as $f_{star} = M_*/M_{bar}$, $f_{H2} = M_{H2}/M_{bar}$, $f_{ HI } = M_{ HI }/M_{bar}$. In our sample of galaxies, we find the range $0.03 \leq f_{star} \leq 0.94 $ for the stellar mass fraction,  $0.01 \leq f_{H2} \leq 0.05 $ for the H2 (molecular gas)
mass fraction, and $0.04 \leq f_{ HI } \leq 0.95$ for the  HI (atomic gas) mass fraction. In Figure [\ref{fig:MsVc-gasfrac}] these fractions are plotted as a function of stellar mass and circular velocity (left and right panels, respectively) of a galaxy. Firstly, we notice that the molecular gas fraction is small in the visible region ($\sim 5\%$) and remains constant as a function of the stellar mass and the circular velocity. Therefore, we argue that at $z\sim 1$ molecular gas does not play a significant role in the kinematics, as occurs in the local SFGs.

Secondly, we observe a gradual decrease (increase) in the mass fraction of the atomic gas (stars) as a function of stellar mass. In particular, at the lower mass end ($\leq 10^{9.5} \mathrm{M_\odot}$) HI dominates, while at higher mass ($\geq 10^{10.2} \ \mathrm{M_\odot}$) stars dominate. At intermediate masses ($ 10^{9.5-10.0} \ \mathrm{M_\odot}$), the fraction of HI and stellar masses is almost 50-50\% (see left panel Figure[\ref{fig:MsVc-gasfrac}]). 

We compare our baryon mass fraction results with local late-type galaxies \citep{Calette2018} shown by solid lines in the left panel of the Figure [\ref{fig:MsVc-gasfrac}]. We note that the objects with $M_* >10^{10} \mathrm{M_\odot }$ tend to have a similar gas and stellar mass fraction as local late-type galaxies, while low-mass $M_* <10^{10} \mathrm{M_\odot }$ objects seem to have a higher HI fraction independent of the local objects. We note that the H2 fraction remains the same. From the comparison of $z\sim 1$ and $z\sim 0$ baryon fractions, a key feature of galaxy evolution emerges, suggesting that the galaxies in our sample are in the middle of their evolutionary path\footnote{ Evolutionary path, a period in which a galaxy gradually accumulates mass and increases in size.}. The brightest (most massive) galaxies appear to be depleting their gas reservoir faster than the faintest galaxies, and because they exhibit this behavior, they appear to be present-day spiral galaxies.

Our goal is to calculate the DM fraction of our galaxies within $R_e, \ R_{opt}$, and $R_{out}$, and therefore we must first determine the baryonic masses within these radii. In our previous work \citep{GS20}, we found that the kinematics of the Q12 sample is similar to local star-forming disk galaxies. This suggests that the radial distribution of stellar and molecular gas masses within these galaxies can be well approximated by the Freeman disk \citep{Freeman}:
\begin{equation}
\label{eq:FMs}
\begin{array}{r@{}l}
M_{*}(<R) = M^{tot}_{*} \ \Big[ 1- \Big( 1+\frac{R}{R_{*}} \Big) \ \exp(\frac{-R}{R_{*}})  \Big]\\ \\
M_{H2}(<R) = M^{tot}_{H2} \ \Big[ 1- \Big( 1+\frac{R}{R_{H2}} \Big) \ \exp(\frac{-R}{R_{H2}}) \Big]\\ \\
M_{HI}(<R) = M^{tot}_{HI} \ \Big[ 1- \Big( 1+\frac{R}{R_{HI}} \Big) \ \exp(\frac{-R}{R_{HI}})  \Big]
\end{array}
\end{equation}
where stars are assumed to be distributed in the stellar disk ($R_* = R_D$) known from photometry, discussed in Section [\ref{sec:data}]. The molecular gas is generally distributed outward through the stellar disk (up to the length of the ionized gas $R_{gas}$); therefore, we take $R_{H2} \equiv R_{gas}$. Here, we estimate the gas scale length $R_{gas}$ by fitting the $H_\alpha$ surface brightness, which is discussed in Appendix[\ref{sec:RH2}]. Moreover, studies of local disk galaxies have shown that the surface brightness of the HI disk is much more extended than that of the H2 disk \citep[][see their Fig. 5]{Jian2010}; see also \cite{Leroy2008} and \citet{Cormier2016}. Therefore, we assume $R_{ HI }=2 \times R_{H2}$, which is a rough estimate, but still reasonable considering that at high-$z$ no information is available on the $M_{ HI }$ (or $M_{H2}$) surface brightness distribution. Thus, the Equation [\ref{eq:FMs}] allows us to estimate $M_{*}$, $M_{ HI }$, and $M_{H2}$ within different radii ($R_e, \ R_{opt}$, and $R_{out}$).


\begin{figure}
        \begin{center}
                \includegraphics[width=\columnwidth]{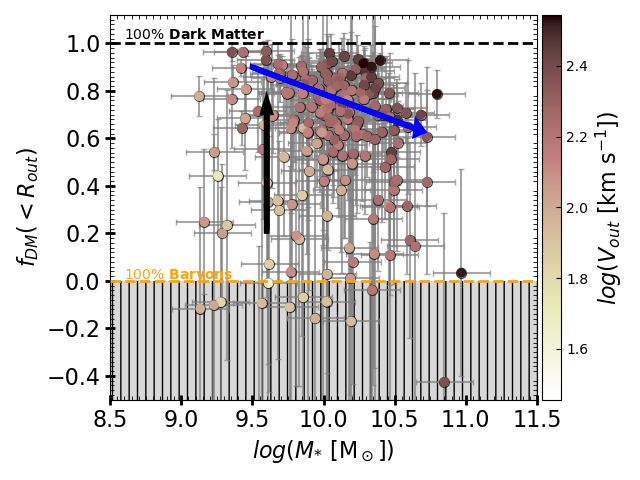}
                \includegraphics[width=\columnwidth]{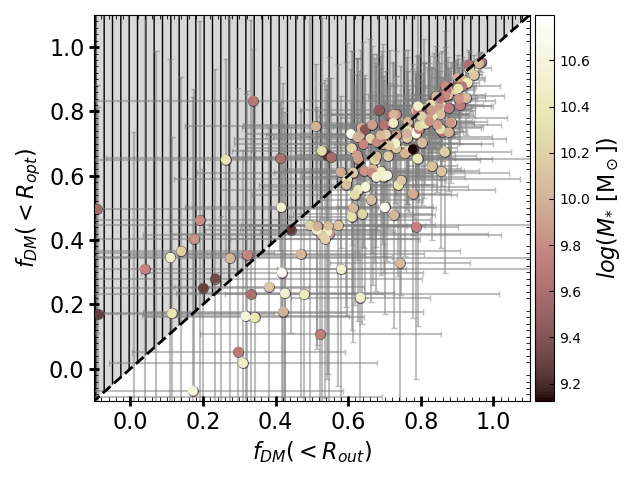}
                \caption{\textbf{DM fraction of individual galaxies.} {\em Upper panel:} Dark matter fraction within $R_{out}$ as a function of stellar mass ($M_*$), color coded by the circular velocity ($V_{out}$) computed at $R_{out}$. The horizontal yellow and back dashed lines show the 100\% baryon and DM regimes, respectively. The gray shaded area shows the forbidden region. The black arrow  indicates that, for a given stellar mass, the fraction of DM in the galaxies increases with increasing circular velocity. The blue arrow  shows a shallow decrease in the DM fraction with increasing stellar mass. {\em Lower panel:} $f_{DM}(<R_{opt})$ vs. $f_{DM}(<R_{out})$, color coded by stellar mass. The gray shaded area represents the  not allowed region (i.e., $f_{DM}(<R_{out}) \nless f_{DM}(<R_{opt}))$. The black dashed line shows the one-to-one relation. For the clarity of the figure (in the lower panel), objects below zero are not shown.}
                \label{fig:fdm1}
        \end{center}
\end{figure} 

\begin{figure}
        \begin{center}
                \includegraphics[width=\columnwidth]{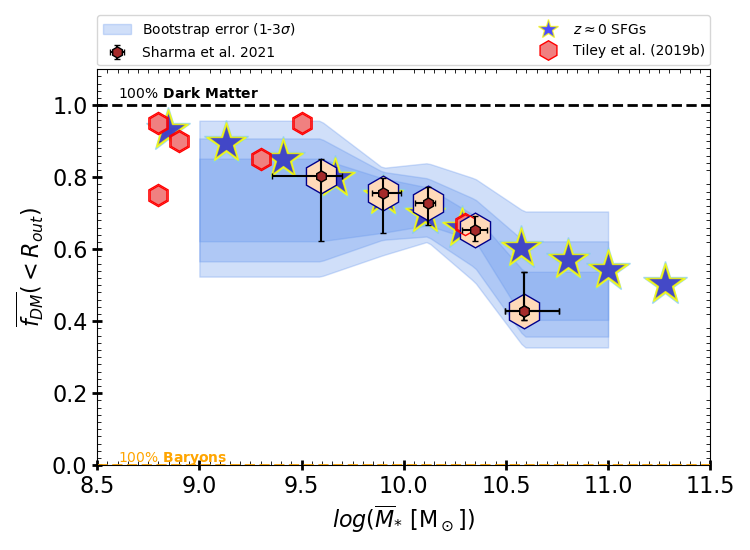}
                \includegraphics[width=\columnwidth]{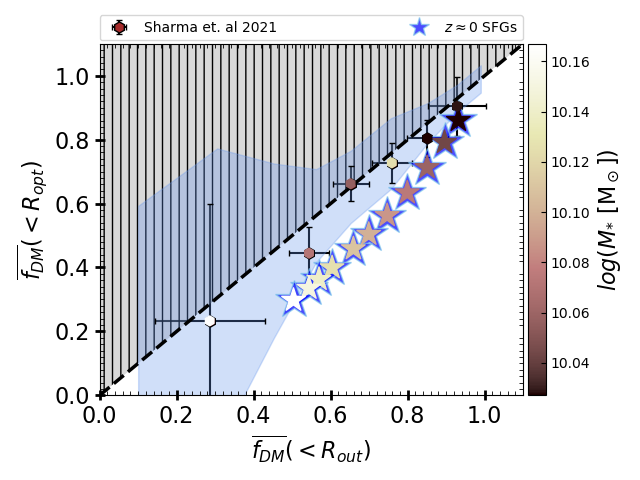}
                \caption{\textbf{DM fraction of ensemble averages.} {\em Upper panel:} Dark matter fraction within $R_{out}$ as a function of stellar mass ($\overline{M}_*$). The brown hexagons with error bars shielded by big light-peach hexagons represent our averaged data, and the blue shaded area shows the 1, 2, and 3$\sigma$ error at DM fraction. For comparison, we show the local star-forming disk galaxies \citep[][]{PS1996} represented by the blue stars (in both panels), where the size of the marker represents the uncertainty in the values.  The red hexagon represents the $z\sim1$ SFGs \citep{AT2019}. {\em Lower panel:} $\overline{f_{DM}}(<R_{opt})$ vs. $\overline{f_{DM}}(<R_{out})$, color coded by stellar mass. The gray shaded area represents the theoretically not allowed region (i.e., $\overline{f_{DM}}(<R_{out}) \nless \overline{f_{DM}}(<R_{opt}))$. The blue shaded area shows the maximum scatter in the relation. The lower panel interior of blue stars are color coded by stellar mass.}
                \label{fig:fdm2}
        \end{center}
\end{figure}


\subsection{Dynamical mass estimates}
\label{sec:Mdyn}
The dynamical mass of a galaxy is defined as 
\begin{equation}
\label{eq:Mdyn}
M_{dyn}(<R) = \kappa(\mathrm{{\tiny R} }) \frac{V^2(R) \ R}{G} 
,\end{equation}
where $V(R)$ is the circular velocity computed at radius $R$ and $\kappa$({\tiny R}) is the geometric factor accounting for the presence of a stellar  disk alongside  the  spherical bulge and  halo (see  \citep{Persic1990}. For $R_{e}$, $R_{opt}$, and $R_{out}$, the values of $\kappa$ are 1.2, 1.05, and 1.0, respectively. We favor a model-independent approach to determine the baryonic, dynamical, and DM masses at a given radius $R$, which differs from the standard RC mass decomposition method, where $M_{bar}+M_{ DM } \simeq M_{dyn}$.

In Figure[\ref{fig:Mdyn-Ms-Mbar}] we show the results of the dynamical mass within $R_{out}$ as a function of stellar and baryonic mass in the left and right panels, respectively. We find that the baryonic masses generally do not exceed the dynamical masses, and a DM component occurs in all objects, as implied independently of their RC profiles. Only a few galaxies are located in the forbidden region; however, they are consistent with $M_{bar}\leq M_{dyn}$ within the $1\sigma$ uncertainty intervals.\footnote{ In this work the stellar, gaseous ($H_2, \ HI$) and dynamical masses were calculated independently using different methods, which have their own errors (and systematic). Therefore, it is possible to encounter the situation where $M_{dyn}< M_{bar}$.}

The bulge mass contribution within 5 kpc is negligible for local spirals; therefore, we do not model it. However, we emphasize that the stellar masses derived from the absolute H-band magnitude (via their M/L ratio) include the bulge mass. Moreover, the geometric parameter $\kappa$({\tiny R}) in the dynamical mass calculation takes into account the distribution of the mass located in the bulge and in the disk.

\section{Results}
\label{sec:results}
Given the information on bayonic and dynamical masses, the DM fractions within radius $R$ can be computed as
\begin{equation}
\label{eq:fdm}
f_{DM} (<R)= 1- \frac{M_{bar} (<R)}{M_{dyn} (<R)}
.\end{equation}
Therefore, using Equation[\ref{eq:fdm}], we computed $f_{ DM }$ within $R_e, \ R_{opt}$, and $ R_{out}$ for all galaxies. We note that owing to the limited spatial resolution in our RCs, the measurements of $f_{DM}$ within $R_e$ are less accurate than for $R_{opt}$ and $R_{out}$. In the following subsection we demonstrate the DM fraction of individual objects as well as in terms of ensemble averages.
\\
\\
\underline{Individual galaxies:} 
In the upper panel of Figure [\ref{fig:fdm1}], the DM fractions within $R_{out}$ are plotted as a function of stellar mass; objects are color coded according to the circular velocity computed at $R_{out}$. First, it is noticeable that for a given stellar mass, the DM fraction seems to increase with increasing circular velocity, whereas the  DM fraction shows a (shallow) decrease with increasing stellar mass. It is important to stress that both the baryonic and dynamical masses are determined with independent methods that have non-negligible errors; this affects the determination of the DM fraction and its errors ($\delta f_{DM}(<R)$). A typical value of  $\delta f_{DM}(<R_{out})$ is $\pm 0.2$; however, at inner radii the uncertainties on the DM fraction reaches 0.4. Second, at $R_{out}$ only $\sim$ 20\% of the objects have $f_{ DM }(R_{out})< 0.5$. Noticeably, the objects with a low DM fraction ($f_{ DM }< 0.5$) are not always the most massive but cover the broad stellar mass range $log(M_* \ [\mathrm{M_\odot}]) \approx 9.2-10.7$, irrespective of the local spirals \citep{PS2019}. Moreover, only 8\% objects fall in the forbidden region. This is very likely due to i) the measurement errors discussed above,  ii) declining RCs, or iii) determination of the RCs themselves. Some examples of such galaxies are shown in the Figure[\ref{fig:exp-RC}].

In the lower panel of Figure [\ref{fig:fdm1}], we investigate the increase in the DM fraction from the optical to the outer radius by plotting $f_{ DM }(R_{opt})$ versus $f_{ DM }(R_{out})$. As expected, 80\% of the objects have $f_{ DM }(R_{out}) >f_{ DM }(R_{opt})$; that is, in the majority of the objects DM fraction increases with radius. However, we note that the perturbed or declining RCs severely affect the 20\% of objects showing $f_{ DM }(R_{out}) <f_{ DM }(R_{opt})$. We will address this problem in our follow-up work, in which we plan to mass model the individual RCs using the component separation method and then determine the DM fraction at various radii. In this work, we proceed by averaging the various DM fractions.
\\
\\
\underline{Ensemble averages :} 
In our work, we focused primarily on the study of individual objects. However, to compare our results with previous studies, we average the DM fraction in a suitable number of stellar mass bins. In particular, we employ affine binning of our 225 objects, keeping 50 objects per bin, and the last bin contains 25 objects. To perform the binning, we use the root mean square statistic (RMS), and the errors are estimated from bootstrap iterations. For each bin, we iterate 500 times, and 50\% samples are taken in each run. Then the errors are assigned for the 68th, 95th, and 99th percentiles (of the scatter). The binning details are further explained and tabulated in Table[\ref{tab:fdm2}]. For simplicity, the averaged quantities are denoted by bar, for example, $\overline{M_*}$ and $\overline{f_{DM}}$. We note that for the stacked data, we plot the $1\sigma$ error on DM fraction and its maximum scatter ($1-3\sigma$) is shown by the blue shaded regions. On the other hand, for stellar mass (being the running variable of the bin) we plot $3\sigma$ errors, which directly shows the maximum deviation from the mean ($\overline{M_*}$).
\begin{table*}
\centering
\begin{tabular}{|l|l|l|l|l|l|l|}
\hline
 & & & & & & \\
Bin & N$_{Obj}$ &  $log(\overline{M}_{*})$ & $\overline{z}$ & $\overline{f_{DM}} (<R_{out})$ & $\overline{f_{DM}} (<R_{opt})$  & $\overline{f_{DM}} (<R_e)$\\
No. & [dpts]& [$M_\odot$] &  & & &   \\[1.0ex]
\hline
\hline
 & & & & & & \\
1 & 50 & 9.59$^{+0.11}_{-0.49}$ & 0.84$^{+0.00}_{-0.01}$ & 0.79$^{+0.05}_{-0.18}$ & 0.73$^{+0.06}_{-0.08}$ & 0.71$^{+0.08}_{-0.11}$  \\[1.2ex]

2 & 50 & 9.89$^{+0.13}_{-0.12}$ & 0.84$^{+0.01}_{-0.01}$ & 0.76$^{+0.01}_{-0.11}$ & 0.70$^{+0.03}_{-0.03}$ & 0.75$^{+0.03}_{-0.00}$  \\[1.2ex]

3 & 50 & 10.12$^{+0.07}_{-0.10}$ & 0.84$^{+0.02}_{-0.03}$ & 0.73$^{+0.01}_{-0.06}$ & 0.66$^{+0.03}_{-0.05}$ & 0.66$^{+0.01}_{-0.05}$  \\[1.2ex]

4 & 50 & 10.35$^{+0.12}_{-0.16}$ & 0.86$^{+0.02}_{-0.04}$ & 0.65$^{+0.04}_{-0.03}$ & 0.65$^{+0.07}_{-0.04}$ & 0.62$^{+0.06}_{-0.11}$  \\[1.2ex]

5 & 27 & 10.59$^{+0.37}_{-0.12}$ & 0.90$^{+0.03}_{-0.02}$ & 0.43$^{+0.11}_{-0.03}$ & 0.30$^{+0.2}_{-0.06}$ & 0.40$^{+0.07}_{-0.21}$  \\[1.2ex]
\hline
\end{tabular} 
\caption{Results of the binned DM fraction. For binning, the objects are first sorted into ascending stellar masses. Then affine binning is performed and 50 objects are allowed per bin. There are a total of 227 objects, so the final bin contains only 27 objects. To perform the binning, RMS statistics and the errors estimated from bootstrap iterations are used. {\em Col. 1:} Bin number; {\em Col. 2:} number of objects (or data points) per bin; {\em Col. 3:} binned stellar masses and their errors showing the minimum and maximum boundaries of each bin, lower and upper bounds respectively; {\em Col. 4:} binned redshift; {\em Cols.-5-7:} binned DM fraction within $R_{out}$, $R_{opt}$, and $R_e$, respectively. \ Here $1\sigma$ errors are given for each data point (except stellar masses). A catalogue of the individual quantities is published along with this table; see Appendix Table [\ref{tab:catalogue}]. }
\label{tab:fdm2}
\end{table*}

In upper panel of Figure[\ref{fig:fdm2}] we show the same results as in the upper panel of Figure[\ref{fig:fdm1}], but for the averaged dataset. From the upper panel, we can clearly see that the DM fraction weakly decreases with increasing stellar mass, and galaxies are DM dominated ($\overline{f_{DM}}(<R_{out})> 50\%$) at any stellar mass value ($log(\overline{M_*} \ [\mathrm{M_\odot}]) \approx 9.5-10.0$). Similarly, in the lower panel of Figure[\ref{fig:fdm2}] we show the same results as in lower panel of Figure[\ref{fig:fdm1}], but for the averaged dataset. In detail, we binned the lower panel of Figure[\ref{fig:fdm1}], in six $f_{DM}(<R_{out})$ bins namely [0.0-0.5, 0.5-0.6, 0.6-0.7, 0.7-0.8, 0.8-0.9, 0.9-1.0], using RMS statistics and errors are $\sqrt{\sigma_i^2/n}$ in each bin.\footnote{where $\sigma_i$ is standard deviation and $n$ is the number of objects per bin.} We can clearly see that the $\overline{f_{ DM }}(R_{out})$ is always larger than $\overline{f_{ DM }}(R_{opt})$; that is, the mass of DM  increases with radius. Both panels of Figure [\ref{fig:fdm2}] show that all of our ensemble averages have $\overline{f_{ DM }} \geq 50 \%$ within $R_{opt}$ till $R_{out}$, except for one data point at higher stellar mass. Nevertheless, ensemble averages of DM fraction in stellar mass bins, shown in Table[\ref{tab:fdm2}], shows that  $\overline{f_{DM}}(<R_{out}) > \overline{f_{DM}}(<R_{opt}) > \overline{f_{DM}}(<R_{e})$, which validates our measurements.

\begin{figure*}
        \begin{center}
                \includegraphics[angle=0,height=6.5truecm,width=16.5truecm]{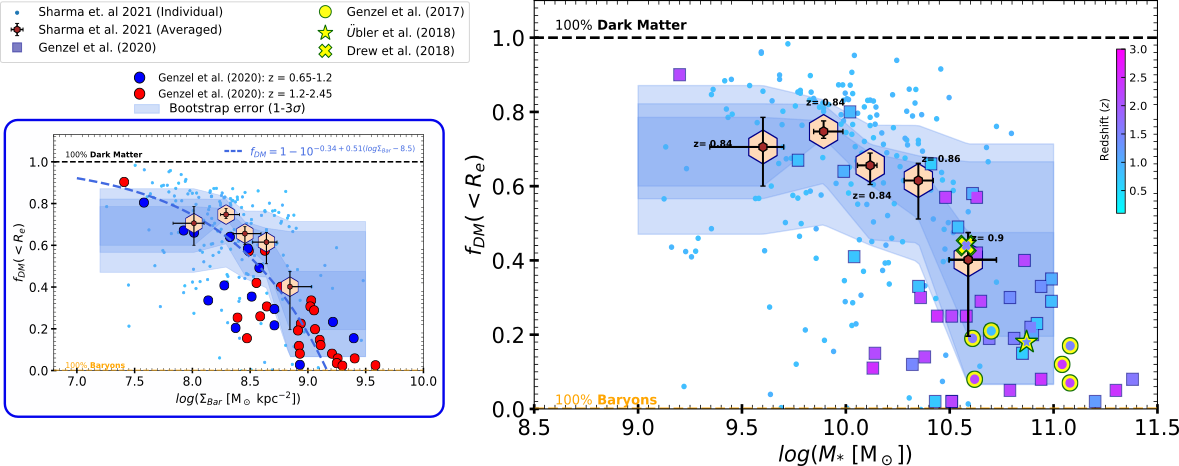}        
                \caption{Comparison of averaged DM fraction (within $R_e$) of our sample with previous high-$z$ studies. {\em Left panel:} Dark matter fraction as a function of baryon surface density $\Sigma_{Bar}(<R_e)= M_{bar} (<R_e)/\pi R_e^2$. The brown hexagons with error bars shielded by big light-peach hexagons represent our averaged dataset, and individual galaxies are shown by small dots, color coded by redshift. The blue shaded area represents the $1-3\sigma$ error on the DM fraction. The dashed blue line shows the relation $f_{DM}(<R_e) = 1- 10^{-0.34+0.51(log\Sigma_{Bar} -8.5)}$ given by \citet{Genzel2020} and their data are plotted in blue ($z= 0.65-1.2$) and red ($z= 1.2-2.45$) filled circles. {\em Right panel:} Dark matter fraction as a function of stellar mass, color coded by redshift. Our dataset and its errors are presented in similar manner in both panels. \citet{Genzel2020} data are shown by filled squares. The measurements of \citet{Genzel2017}, \citet{Drew2018}, and \citet{Ubler2018} are shown by yellow circles, a star and crosses, respectively, where the interior of each marker is color coded by redshift. The legend of the right plot is shown in the upper of left plot. }
                \label{fig:fdm3}
        \end{center}
\end{figure*}

\section{Discussion} 
\label{sec:discussion}
In \citet{GS20} we showed that the outer RCs of $z\sim 1$ SFGs (having $v/\sigma >1$) are similar to those of local ($z\approx 0$) star-forming disk galaxies \citep{PS1996}. In this work, we intend to compare the DM fraction of $z\sim 1$ SFGs ($v/\sigma >1$) with those of $z\approx 0$ star-forming disks. The results are shown in Figure [\ref{fig:fdm2}]. In the upper panel, we note that SFGs at $z\sim 1$ and $z\approx 0$ follow a similar trend in the $\overline{f_{ DM }}-\overline{M}_{*}$ plane; these SFGs also seem to have almost the same fraction of DM in the outer radius. In the lower panel of Figure [\ref{fig:fdm2}], we again observed a similar DM behavior as in the locals, but with a relatively higher DM fraction inside $R_{opt}$, which suggest that the radial profile of dark and luminous matter has evolved in the past 7 Gyrs. This could be because at $z\sim 1$ the DM halo has already formed, while the stellar disk inside $R_{opt}$ is still in the process of formation.

Our results are in fair agreement with those of \citet{AT2019} of similar SFGs. In particular, their sample lies in the redshift range $0.6 \lesssim z \lesssim 2.2$; therefore, we narrowed down their sample to the galaxies that lie between $0.7 \lesssim z \lesssim 1.0$ and over-plotted them in Figure[\ref{fig:fdm2}]. As shown in the plot, there is a good agreement between the two studies. We note that \cite{AT2019} computed $\overline{f_{ DM }}$ within $6 \ R_D$, while we compute within $R_{out}=5 \ R_D$. 

Next, we compared our work with \citet{Genzel2017, Drew2018}, and \citet{Ubler2018}; see right panel Figure[\ref{fig:fdm3}]. In these studies, $f_{DM}$ is computed within $R_e$, where our RCs are not spatially resolved and thus $f_{DM}(<R_e)$ is subject to larger uncertainties than $f_{DM}$ within $R_{opt}$ and $R_{out}$. Even so our four massive ($log(M_*)\sim 10.5-10.9$) galaxies with $f_{DM}(R_{out})<0.2$ are very close to the results of \citet{Genzel2017} and \citet{Ubler2018}. Moreover, we agree closely with \citet{Drew2018}. We also compare our work with the recent work of \citet{Genzel2020} (squares color coded for redshift). Their sample has seven objects with $9.0 \lesssim \log (M_*) \lesssim 10.5$ within the redshift range of $0.5 <z \lesssim 1.0$ and have a DM fraction $\geq 0.5$; these values agree well with our individual DM fraction within $R_e$. However, the ensemble mean of DM fraction within $R_e$ is always greater than 40\%, showing dissimilarity with the majority of \citet{Genzel2020} objects. In particular, the mean of the massive bins for which ([$\log (\overline{M}_*), \overline{f_{ DM }}, \overline{z}] = [10.6, 0.53, 0.9] $) shows a much higher DM fraction than \citet{Genzel2020}. We would like to emphasize that the majority of their sample\footnote{We compare SED driven stellar masses of the \citet{Genzel2020} sample.} have high stellar masses ($log(M_*)> 10.5$) and a low DM fraction ($f_{ DM }< 35\%$), which is not the case for our sample. 

Moreover, we also compared our results with the $f_{DM}(<R_e)$ versus $\Sigma_{Bar}(<R_e) $ relation put forward by \citet{Genzel2020} (see left panel Figure[\ref{fig:fdm3}]), where $\Sigma_{Bar}(<R) = M_{bar}(<R)/\pi R^2 $. This relationship can be pivotal for both the DM nature and evolution of disk galaxies. We note that for low surface brightness ($log(\Sigma_{Bar}(<R_e) \ \mathrm{M_\odot \ kpc^{-2}}) < 8.7 $) galaxies, their relation is in fair agreement with our data, but at high $\Sigma_{Bar}$ it underestimates DM fractions. We suspect that the high bulge-to-total (B/T) disk ratio used in \citet{Genzel2020} artificially reduces the DM within the effective radius.

Furthermore, while the DM -fractions at $z\sim 1$ and $z\sim 0$ are roughly the same, the scatter in the $f_{ DM }-M_{*}$ relation suggests that these galaxies are still in the process of building (or acclimating) the distribution of baryons and DM; that is, they are at different stages of their disk formation. However, it could also be due to the diversity in the DM halo properties (e.g., concentration and core radius or density), which are closely coupled to the properties of the baryonic matter (e.g., halo spin parameter and baryon fraction).

\section{Summary} 
\label{sec:summary}
In this work, we have studied the fraction of DM in $z\sim 1$ SFGs. We used our previous study of RCs \citep{GS20} to compute accurate dynamical masses. In particular, we exploit 225 high-quality galaxies from the Q12 sample studied in \citet{GS20}. This sample lies between redshift $0.7 \lesssim z \lesssim 1.0$ with stellar masses $log(M_{*} \ \mathrm{[M_\odot]}) = 9.0-11.0$ and circular velocities $log(V_{out} \ \mathrm{km \ s^{-1}})= 1.45-2.83$. We estimate the total baryonic masses using scaling relations of atomic and molecular gas masses given by \citet{Lagos2011} and \citet{Tacconi2018} respectively (see  Section [\ref{sec:MHI}, \ref{sec:MH2} \& \ref{sec:Mbar}]). Subsequently, the DM fractions were computed using dynamical mass estimates from RCs. In Section [\ref{sec:results}], we showed that only $\sim 6\%$ of objects have low DM fractions ($0.0<f_{ DM }\leq 0.2$), but these objects are not necessarily massive. That is, a low DM fraction can also be found in low-mass galaxies ($log(M_{*} \ \mathrm{[M_\odot]}) <9.5 $); however, uncertainties are higher there. Nonetheless, the majority ($\gtrsim 72\%$) of our sample contains DM-dominated ($f_{ DM } \gtrsim 0.5-0.99$) disks with a median radius $\sim 9$ kpc (see Figure [\ref{fig:fdm1}], [\ref{fig:fdm2}] and [\ref{fig:fdm3}]). Our results are in agreement with a previous high-$z$ study of the DM fraction of \citet{AT2019} and are also consistent with local star-forming disk galaxies \citep[][]{PS1996}. We conclude that

\begin{enumerate}
\item The majority of star-forming disk-like galaxies ($v/\sigma >1$) at $0.7\lesssim z \lesssim 1.0$ have outer ($\approx 5
- 9$ kpc) disks dominated by DM  (see Figure[\ref{fig:fdm1} \& \ref{fig:fdm2}]).

\item Baryon-dominated galaxies exist at high-$z$, similar to the local Universe, but they are very few ($20 \%$).

\item Star-forming galaxies at $z \sim 1$ have similar or slightly higher ($\sim 20\%$) DM fractions than the local ones within $R_{opt}$ (see Figure[\ref{fig:fdm2}]). 

\item The scattering in the DM fractions at given stellar masses and circular velocities is larger at $z\sim 1$ than at $z\sim 0$. We interpret this as a consequence of ongoing galaxy formation or processes of evolution.
\end{enumerate}
To gain a more fundamental understanding of the results presented in this work, as well as our earlier findings, we are working on the mass decomposition of RCs (Sharma et al. in preparation), which will shed light on the structural properties of DM and its interplay with baryons.

\begin{acknowledgements}
We thank the anonymous referee for their constructive comments and suggestions, which have significantly improved the quality of the manuscript. We thank A. Tiley for providing us the SED driven stellar masses of KROSS sample. We thank Andrea Lapi for his comments. G.S. thanks M. Petac for his fruitful discussion and various comments in the entire period of this work. GvdV acknowledges funding from the European Research Council (ERC) under the European Union’s Horizon 2020 research and innovation programme under grant agreement No 724857 (Consolidator GrantArcheoDyn).
\end{acknowledgements}

%
\bibliographystyle{aa} 
\bibliography{DM_fraction} 
%

%

\begin{appendix} 
\section{SED-driven stellar masses}
\begin{figure}
        \begin{center}
                \includegraphics[width=\columnwidth]{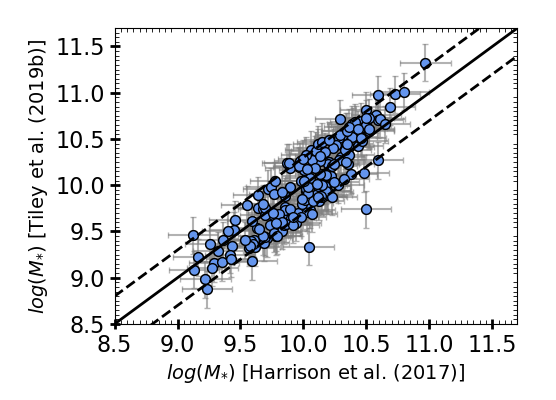}
                \caption{Comparison of stellar masses derived from SED fitting \citep{AT2019} and fixed mass-to-light ratio \citep{H17}. The black solid and dashed line shows the one-to-one relation and $0.2$ dex scatter, respectively.}
                \label{fig:Ms-comparison}
        \end{center}
\end{figure}

We discuss the stellar masses derived from two different approaches: SED fitting techniques \citep{AT2019} and fixed mass-to-light ratio stellar masses \citep{H17}. In particular, \citet{H17} uses H-band absolute magnitude to derive the fixed mass-to-light stellar masses; details are given in Section[\ref{sec:Mstar}]. On the other hand, \citet{AT2019} stellar masses are derived using the {\sc Le Phare} \citep{Arnouts1999, Ilbert2006} SED fitting tool. The {\sc Le Phare} compares the suites of modeled SED of objects from the observed SED. Where the observed SED of our sample is derived from optical and NIR photometric bands (U, B, V, R, I, J, H, and K), in some cases we also used the IRAC mid-infrared bands. In modeling, the stellar population synthesis model is derived from \citet{Bruzual2003}, and stellar masses are calculated using \citet{Chabrier_2003} initial mass function (IMF). The {\sc Le Phare} routine fits the extinction, metallicity, age, star formation, and stellar masses and allows for a single burst, exponential decline, and constant star formation histories. Details of stellar-mass computation are available in \citet{AT2019}. In Figure[\ref{fig:Ms-comparison}], we show that the \citet{AT2019} and \citet{H17} stellar masses are in full agreement with an intrinsic scatter of $0.2$ dex.  \citet{AT2019} have not yet published the stellar masses of full KROSS sample, but we have access to them (in private communication); therefore, we are providing a consistency check.

\section{Star formation rates} 
\label{sec:SFR}
In general, SFRs in the ultra-violet, optical, or NIR (0.1-5 $\mu m$) probe the direct star light of galaxies, while SFRs in mid- or far-infrared  (5-1000$\mu m$) probe the stellar light reprocessed by dust. Therefore, the total SFR of any galaxy can be obtained by the linear combination of stellar and IR luminosities \citep{Kennicutt1998, Calzetti2007, Kennicutt2012}, for example,

\begin{equation}
\label{eq:SFR-K98}
\mathrm{SFR} \ [\mathrm{M_{\odot} \ yr^{-1}}] \ =  \ C_{H\alpha}  L_{H\alpha} \ + \  C_{24\mu m}  L_{24 \mu m}
,\end{equation}
where $C_{H\alpha}$ and $C_{24\mu m}$ are the calibration constants, depends on star formation histories and IMFs. We can also derive the total SFR solely from stellar light (e.g., L$_{H\alpha}$), but then it requires an additional dust attenuation (or extinction) correction, which is defined as $\mathrm{SFR}^{tot} = \mathrm{SFR}^{obs} \times 10^{A_{\nu}}$, where $A_{\nu}$ is the attenuation correction at given wavelength. 

\begin{figure}
        \begin{center}
                \includegraphics[width=\columnwidth]{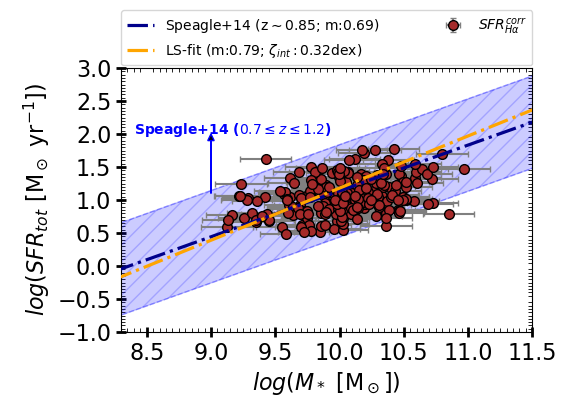}
                \caption{Star formation rate (SFR) vs. total stellar mass ($M_*$), tracing the MS of galaxies, where $SFR_{tot} =  SFR_{H\alpha}$ corrected for dust reddening. The brown filled circles represent the data used in this work. The dashed-dotted orange line shows the least-squares fit to our data, and the blue line shows the \citet{speagle14} MS relation at $z\sim 0.85$. The shaded blue region represents the MS limit between redshift $ 0.7 \leq z \leq 1.2$, which includes 0.3 dex uncertainty at each redshift. }
                \label{fig:SFR}
        \end{center}
\end{figure}

In our sample we do not have possibility to trace the luminosities from UV to the mid- or far-infrared. However, we have access to $H_{\alpha}$, H-, and K-band luminosities. Therefore, we estimated the SFR from $H_{\alpha}$ using  \citet{Kennicutt1998} relation given as
\begin{equation}
\label{eq:SFR-Ha}
SFR_{H\alpha} \ [\mathrm{M_{\odot} \ yr^{-1}}] \ = 4.677 \times 10^{-42} \ L_{H\alpha} \ [\mathrm{erg \ s^{-1}}]
,\end{equation}
where calibration constants, $C_{H\alpha} \ = \ 4.677 \times 10^{-42} \  M_\odot yr^{-1} erg^{-1} s$, include the correction factor of 1.7 due to the \citet[][]{Chabrier_2003} IMF and \citet{Salpeter1955} IMF used by \citet{Kennicutt1998}. Furthermore, we correct for dust reddening (extinction), based on the average value ($A_{Vgas}=1.43$) derived in \citet{stott2016} using the SED fitting. The  $A_{Vgas}=1.43$ is corrected for stars using $A_{Vgas}/A_{Vstar} \approx 1.7$ \citep{Calzetti2000, Pannella2015}. In the Figure[\ref{fig:SFR}], we show the results and compare them with SFR-$M_*$ relation given by \citet{speagle14}, MS of SFGs. In particular, we derive the \citet{speagle14} MS relation between $0.7 \leq z \leq 1.2$ including $0.3 \  dex$ error (shown by blue shaded area), and the best fit is given for $z=0.85,$ which has a slope 0.69. For our sample we derive the best fit using the least-squares method, which includes the errors in both axis. We obtain a slope of  $0.79\pm 0.1$ with intrinsic scatter $0.32 \  dex$. Using a similar approach for the SFR determination, \citet{stott2016} also observed a similar scatter in the SFR-M$_*$ relation, which assures our measurements.

\section{Gas disk length}
\label{sec:RH2}
\begin{figure}[h]
  \begin{center}
    \includegraphics[angle=0,height=6.0truecm,width=8.0truecm]{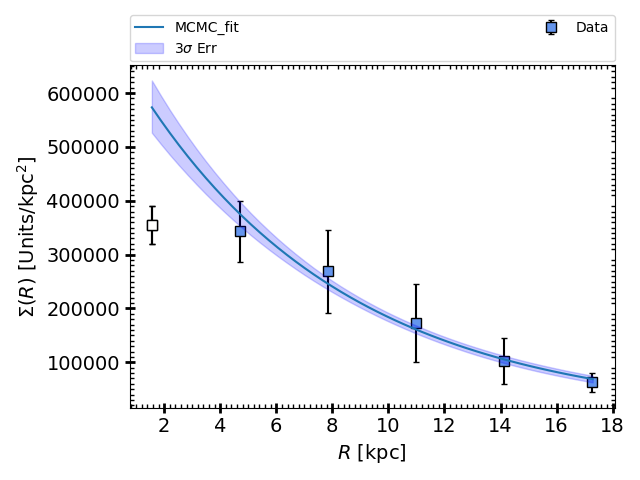}
    \includegraphics[angle=0,height=6.5truecm,width=6.5truecm]{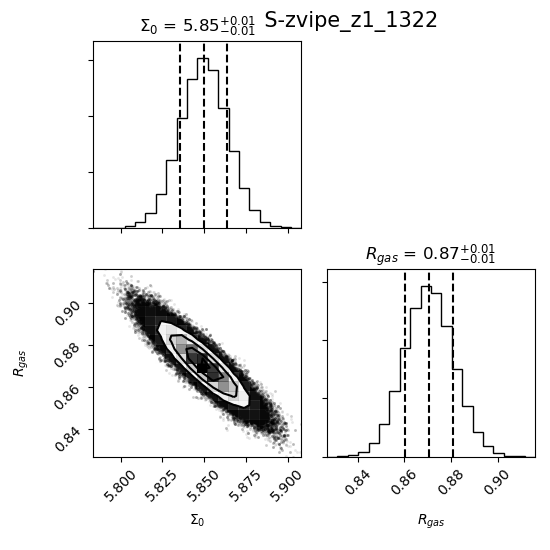}
    \caption{{\em Upper panel:} Observed surface brightness of $H_\alpha$ gas. The MCMC fit is shown in blue accompanied by $1\sigma$ error (blue shaded area). The first point (open square) is not used in fitting owing to the limitation assigned by Barolo (discussed in \citealt{GS20}). The value $\Sigma_0$ keeps the dimension $\mathrm{units/kpc^2}$, where units $= \mathrm{ ergs^{1} cm^{-2} \mu^{-11} \ 1e+17 \ km/s}$. {\em Lower panel:} Posterior distribution of $\Sigma_0$ and $R_{gas}$, where vertical dashed lines show the 16, 50, 84 percentiles from left to right, respectively. }
    \label{fig:SH2}
  \end{center}
\end{figure}

We assume molecular and atomic gas has a spatial distribution described by an exponential profile
\begin{equation}
\label{eq:SH2}
\Sigma_{gas} = \Sigma_0 \exp(\frac{-R}{R_{gas}})
,\end{equation}
where $\Sigma_0$ is central surface density and $R_{gas}$ is gaseous disk scale length. To estimate the value of $R_{gas}$, we fit the observed surface density of ionized gas\footnote{In principle star formation is embedded in the molecular gas clouds; therefore, surface density of ionized gas can be used to trace the gas distribution.} following Equation[\ref{eq:SH2}] using Markov Chain Monte Carlo (MCMC) sampling. During the fitting procedure $\Sigma_0$ and $R_{gas}$ are the free parameter defined in the range $1 \leq log(\Sigma_0) \leq 9$ and $-0.5 \leq log(R_{gas} \ [\mathrm{kpc}]) \leq 1.77$, respectively. An example of best fit and posterior distribution is shown in Figure[\ref{fig:SH2}].

\section{Extra}
\label{sec:extra-fig}
\begin{figure*}
        \begin{center}
                \includegraphics[angle=0,height=3.5truecm,width=4.5truecm]{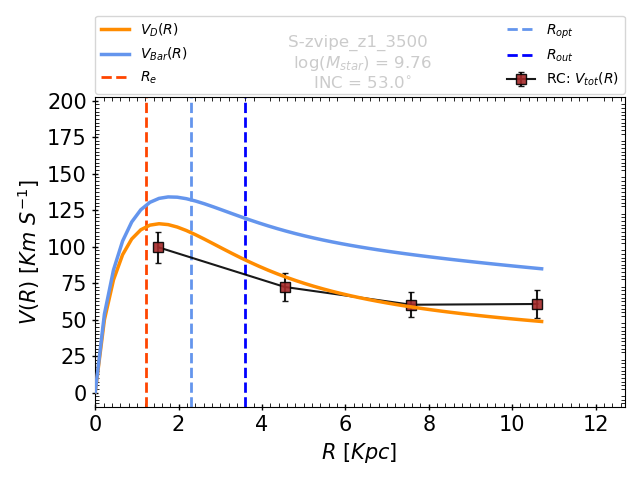}
\includegraphics[angle=0,height=3.5truecm,width=4.5truecm]{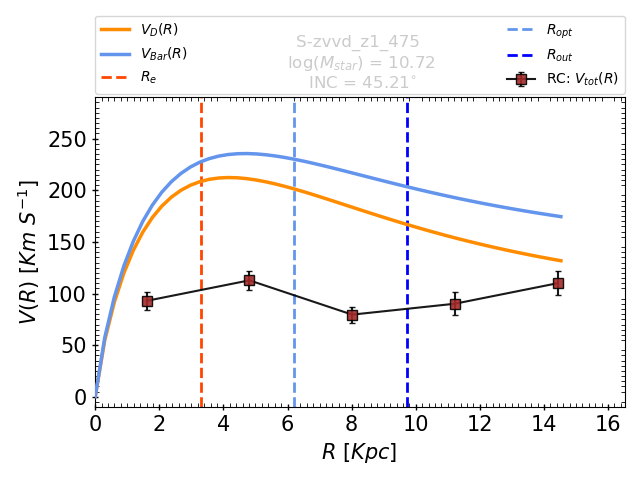}
\includegraphics[angle=0,height=3.5truecm,width=4.5truecm]{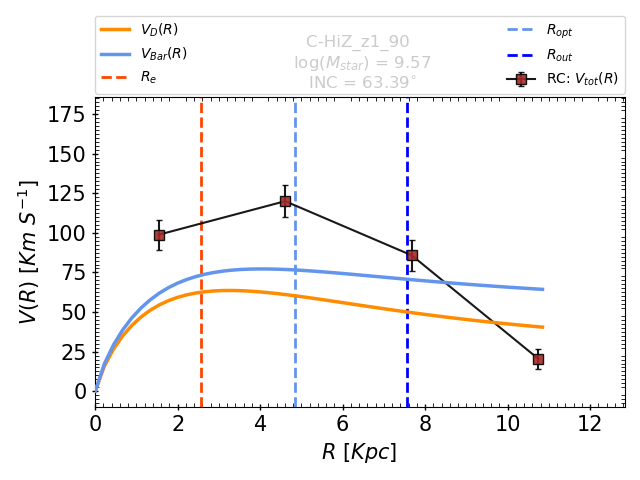}
\includegraphics[angle=0,height=3.5truecm,width=4.5truecm]{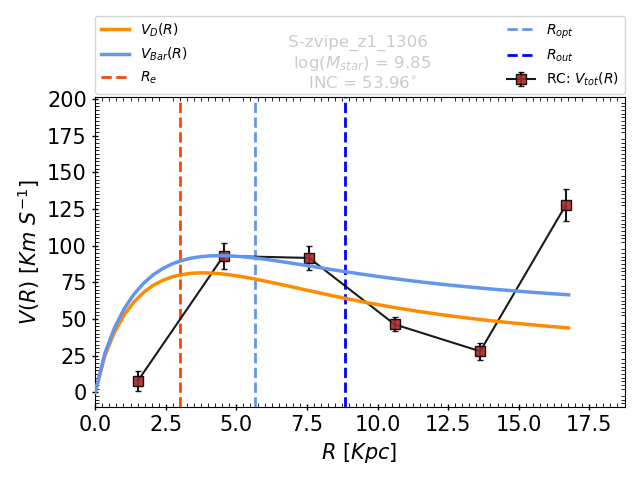}              
                \caption{Few example RCs for which either dynamical masses are unexpected or are a DM fraction. The color codes are as follows: Observed RC is shown by brown squares with the error bars connected with a black line; the orange and blue curves indicate the stellar and baryonic mass velocity curves derived assuming \citet{Freeman} disk for given total $M_*$ and $M_{bar}$. The vertical red, light blue, and dashed blue lines represent the $R_e$, $R_{opt}$, and $R_{out}$ respectively. {\em From left to right:} First and second type of RCs seem to have either overestimated baryonic mass or wrong RC, which leads to smaller dynamical masses than baronic masses, and thus a negative DM fraction. The third type of RCs are announced declining, thus giving us $f_{DM}(<R_{opt}) > f_{DM}(<R_{out})$, which is generally not seen in the local SFGs. The fourth type of RCs are disturbed, thus discarded from the analysis.}
                \label{fig:exp-RC}
        \end{center}
\end{figure*}

Here, we provide examples of RCs which are providing wrong dynamical mass estimates either because of problem in the RC or in the determination of baryonic mass (see Figure[\ref{fig:exp-RC}]).

\begin{table*}[h]
        {\footnotesize
                {\centerline {\sc Descriptions of Columns}}
                \begin{tabularx}{\textwidth}{lccl}
                        \hline
                        Number & Name & Units & Description \\
                        \hline
                        \hline
                        1 & KID & & KROSS ID. \citep[][hereafter Ref: H17]{H17}.\\
                        
                        2 & Name && Object name (Ref: H17).\\   
                        
                        3 & Redshift && Redshift from H$\alpha$ (Ref: H17).\\
                        
                        4 & INC & degrees &Inclination angle (Ref: H17).\\
                        
                        5--6 & Re, Re\_err &$\mathrm{kpc}$& Deconvolved continuum half-light radii, $R_{1/2}$, from the
                        image and error (Ref: H17).\\
                        
                        7--8 & Ve, Ve\_err &$\mathrm{km \ s^{-1}}$& Rotation velocity computed at $\ R_e$ and error.\\
                        
                        9--10 & Vopt, Vopt\_err &$\mathrm{km \ s^{-1}}$& Rotation velocity computed at $\ 1.89 \ R_e$ or $3.2 \ R_D$ (where $R_D = 0.59 \ R_e$) and error.\\

                        11--12 & Vout, Vout\_err &$\mathrm{km \ s^{-1}}$& Rotation velocity computed at $\ 2.95 \ R_e$ or $5 \ R_D$ and error.\\
                        
                        13--14 & Mstar, Mstar\_err &$\mathrm{M_\odot}$& Stellar mass derived from fixed mass-to-light ratio and error.\\
                        
                        15--16 & MH2, MH2\_err &$\mathrm{M_\odot}$& Molecular gas mass derived from derived from scaling relation and error.\\
                        
                        17--18 & MHI, MHI\_err &$\mathrm{M_\odot}$& Atomic gas mass (includes helium content) derived from derived from scaling relation and error.\\
                        
                        19 & Rot\_Disp\_ratio\_int & -- & Rotation to dispersion ratio derived from intrinsic velocity (inclination corrected) \\
                        
                        20--21 & Rgas, Rgas\_err &$\log(\mathrm{kpc})$& Gaseous disk length derived from $H_\alpha$ surface brightness and error\\
                        
                        22 & Rout\_Flag & -- & If `F' then $V_{out}$ cannot be extrapolated due to large difference between last point in \\
                           &            &     & rotation curve and $R_{out}$.\\
                        
                        23 & Mstar\_flag & -- & If `F' then $M_{dyn}<M_{*}$.\\
                        24 & Wiggle\_Flag & -- & If `F' then RC is badly perturbed.\\
                        25 & Rgas\_flag & -- & If `F' then Rgas is unable to determine.\\
                        
                        \hline
                \end{tabularx}
        }
        \caption{\label{tab:catalogue}
                Details of the columns provided in the catalogue associated with this paper. For details of other relevant physical quantities such as velocity dispersion, $H_\alpha$ luminosity, see the \href{https://doi.org/10.7910/DVN/MHRG4O}{catalogue} released with \citet{GS20}.
        }
\end{table*}

With this paper, we release a catalogue of 225 SFGs studied in this work. In Table[\ref{tab:catalogue}], we describe the
columns of the catalogue.

\end{appendix}

\end{document}